\documentclass[preprint,
 amsmath,amssymb,
 aps,
]{revtex4-2}

\usepackage{graphicx}
\usepackage{dcolumn}
\usepackage{bm}

\usepackage[utf8]{inputenc}
\usepackage[T1]{fontenc}
\usepackage{braket}
\newcommand{\bracket}[1]{\langle #1 \rangle}

\begin{document}


\title{Improving quantum-battery charging\\ via unidirectional quantum jumps to metastable state}

\author{Ewelina Lange}
 \email{ewelan@amu.edu.pl}
\affiliation{Institute of Spintronics and Quantum Information, Faculty of Physics and Astronomy, Adam Mickiewicz University, 61-614 Pozna\'{n}, Poland}

\author{Grzegorz Chimczak}
\affiliation{Institute of Spintronics and Quantum Information, Faculty of Physics and Astronomy, Adam Mickiewicz University, 61-614 Pozna\'{n}, Poland}


\author{Anna Kowalewska-Kud{\l}aszyk}
\affiliation{Institute of Spintronics and Quantum Information, Faculty of Physics and Astronomy, Adam Mickiewicz University, 61-614 Pozna\'{n}, Poland}

\author{Kishore Thapliyal}
\affiliation{Joint Laboratory of Optics of Palack\'{y} University
and Institute of Physics of the Czech Academy of Sciences,
Institute of Physics of the Czech Academy of Sciences, 17.
listopadu 1154/50a, 779 00 Olomouc, Czech Republic}
\affiliation{Department of Physics, University of Oslo, 0316 Oslo, Norway}

\author{Jan Pe\v{r}ina Jr.}
\affiliation{Joint Laboratory of Optics of Palack\'{y} University
and Institute of Physics of the Czech Academy of Sciences,
Institute of Physics of the Czech Academy of Sciences, 17.
listopadu 1154/50a, 779 00 Olomouc, Czech Republic}


\date{\today}

\begin{abstract}
In the process of charging a quantum battery, quantum jumps play a detrimental role because they induce transitions from higher to lower energy levels, leading to energy dissipation. This behavior directly opposes the fundamental objective of the charging protocol, which is to increase the population of higher-energy states and store usable energy in the system. Furthermore, decoherence introduced by randomness of quantum jumps leads to a mixed state, thereby reducing the amount of energy that can be extracted from the charged quantum battery. Here, we propose a quantum battery comprising an ensemble of three-level atoms in the $\Lambda$ configuration to store energy in the metastable state over a long time. We demonstrate that, when spontaneous atomic transitions are restricted to occur solely from the excited state to a metastable state, with decay to the ground state completely suppressed, quantum jumps acquire a constructive character. Under these conditions, they drive the quantum battery to a fully charged state, render the entire stored energy extractable, and shorten the duration of the second stage of the charging protocol.
\end{abstract}

\maketitle


\section*{Introduction}

A quantum battery (QB) is usually defined as a $d$-dimensional quantum system, which is able to store energy. This means that a quantum system can be considered as a quantum battery when the difference between the largest and the smallest eigenvalue of its Hamiltonian $H_B$ is greater than zero~\cite{AlickiPRE13,Colloquium}. A QB may be charged through either a direct or an indirect charging process. In the direct charging method, the energy is delivered by a classical laser field during the charging period $\tau$. In the indirect charging method, the energy is transferred to the quantum battery from an auxiliary quantum system, described by the Hamiltonian $H_C$, which is referred to as the charger. An example of charging a quantum battery with a charger is the very well-known system based on the Dicke model~\cite{FerraroPRL18}. The cavity prepared initially in the Fock state $|n\rangle$ plays the role of a charger, and an ensemble of $n$ two-level atoms plays the role of a quantum battery. More recently, quantum batteries are receiving growing interest across a wide range of scientific fields such as mesoscopic physics, quantum information theory, or many-body physics because it is possible to utilise quantum phenomena to enhance the functionality and performance of these energy storage devices. Collective many-body interactions and quantum correlations between subsystems can lead to a superextensive scaling of the charging power. This means that the charging time of a quantum battery improved by these phenomena shortens with its capacity. In the case of the Dicke quantum battery, the charging speedup is a result of the collective interaction of all the atoms with the quantised cavity mode~\cite{LewensteinPRR20} and the collective charging power scales as $n^{3/2}$ in this type of QB~\cite{FerraroPRL18}. Fast charging of quantum batteries is the primary motivation for many researchers to study their properties and pursue experimental realisations. Unfortunately, significant challenges remain in designing a practical quantum battery that can be realised experimentally. To the best of our knowledge, there is only one experimental demonstration of Dicke QB consisting of many atoms ($n>1$)~\cite{QuachSciAdv22}. The primary challenge stems from the fact that all experimental systems are inherently open, allowing energy to leak into the surrounding environment. This poses a significant problem for quantum batteries as their fundamental purpose is to store energy efficiently and reliably over long timescales, without loss, so that it remains fully available when needed. Moreover, interaction between the system and the environment introduces decoherence that diminishes the quantum correlations, which are beneficial for speedup of the charging process. It is not only the speed of the charging process that suffers from decoherence. Individual decay processes of two-level atoms also lead to the aging of QB, i.e., QB can store less energy in subsequent charging processes~\cite{AgingQB}. Therefore, many theoretical studies have been devoted to finding ways to eliminate the unwanted effects of decoherence on the energy storage and on the charging process in QBs. There are different strategies to suppress the unwanted effects of decoherence. In order to prevent the decrease of energy in QB during storage, Gherardini \emph{et al.}~\cite{GherardiniPRR20} proposed using the quantum Zeno effect, i.e., they designed a protocol that contains sequential projective measurements of energy. Other researchers proposed QB schemes, which exploit a decoherence-free subspace~\cite{LiuJPCC19,SantosPRE19,QuachPRApp20,Hu_2022}. Yet another strategy to reduce the undesirable self-discharging of QB is quantum reservoir engineering, which exploits the memory effects present in non-Markovian dynamics~\cite{Kamin_2020}. Interestingly, dissipation can also play a positive role in some aspects of the charging process. Dissipation can be used, for example, to achieve a large enhancement in charging power~\cite{MayoPRA22,PokhrelPRL25} or to perform a directed nonreciprocal energy flow from the quantum charger to the QB~\cite{AhmadiPRL24}. Although such advantages achieved using dissipation are impressive, one should keep in mind that they come at the cost of charging efficiency, i.e. some fraction of energy from an energy source is lost to the environment. 

Building upon these developments, recent research has explored how to utilize environmental and ancillary degrees of freedom to not only mitigate losses but to fundamentally unlock higher energy capacities. For instance, the framework of superoptimal charging has been introduced, \cite{AhmadiPRApp25}, demonstrating that by replacing standard coherent interactions with a dissipative interaction mediated by an engineered shared reservoir, the restriction on energy transfer usually imposed by local damping can be overcome. This method makes use of collective effects from the environment to facilitate a beneficial redistribution of energy, allowing the battery to accumulate near-unlimited energy while maintaining zero optimal detuning, which significantly simplifies the experimental implementation compared to conventional optimal protocols.

Parallel to environmental engineering, the use of ancillary catalyst systems—such as qubits or harmonic oscillators prepared in the ground state—has proven effective in boosting energy transfer without the catalyst itself storing work or becoming significantly entangled with the battery \cite{RodriguezPRA23}. This catalytic approach specifically addresses the problem where the laser drive frequency is out of tune with the global supermodes of the interacting charger-battery system. By mediating the interaction, the catalyst stabilizes the charging process against frequency fluctuations and eliminates the need for precise probing of coupling strengths, thereby ensuring an efficient and robust energy accumulation even in the presence of noise. Together, these proposals suggest that the strategic use of dissipation and mediated interactions can transform the environment from a source of decoherence into a useful tool for optimizing quantum storage devices

In this paper, we investigate the potential of utilizing quantum jumps—the dissipative processes that typically cause decoherence—to improve the charging dynamics of a specially designed QB, one that is capable of reliably storing energy over extended periods. To this end, we propose as a QB an ensemble of three-level atoms in the $\Lambda$ configuration and their long-living metastable levels as perfect energy storage.
A QB comprising a single $\Lambda$-type atom charged by crossing many thermal cavities was previously investigated by Bele\~no \emph{et al.}~\cite{BelenoNJP24}. To our knowledge, this is the only study that exploits a $\Lambda$-type system for the long-term storage of energy, which is particularly striking given that such systems have been extensively employed for many years in a variety of proposals aimed at the reliable preservation of quantum information (for example, see~\cite{ChimczakPRA05,ChimczakPRA05Eff,ChimczakPRA07,ChimczakPRA08,ChimczakPRA09}). In our proposal, we introduce an unconventional quantum $\Lambda$-type system in which quantum jumps are possible only from the excited state to the metastable state but not to the ground state. We show that these \emph{unidirectional quantum jumps} lead to a steady state of the considered system, which is a pure state. This is a crucial feature of the system under investigation, as it ensures that all energy stored in the quantum battery is fully extractable, i.e., ergotropy~\cite{AllahverdyanEPL04} is equal to the energy stored in the QB. We also show that such unidirectional quantum jumps can significantly decrease the time of the second stage of the charging protocol in the weak coupling regime, and because of them, one can fully charge the QB. 
Quantum jumps generally play an important role in the dynamics of the system and, for example, can even move the location of an exceptional point in the parameter space~\cite{LEP,chimczak_effect}, but usually this is an adverse effect due to its randomness. Here, we show that unidirectional quantum jumps in the unconventional $\Lambda$-type system play a constructive role. Finally, we prove that a ${}^{87}\mathrm{Rb}$ atom can effectively work as the unconventional $\Lambda$-type system in a two-stage charging protocol.

\section*{Results}

\subsection*{The model}

\begin{figure}[ht]
\centering
\includegraphics[width=0.9\linewidth]{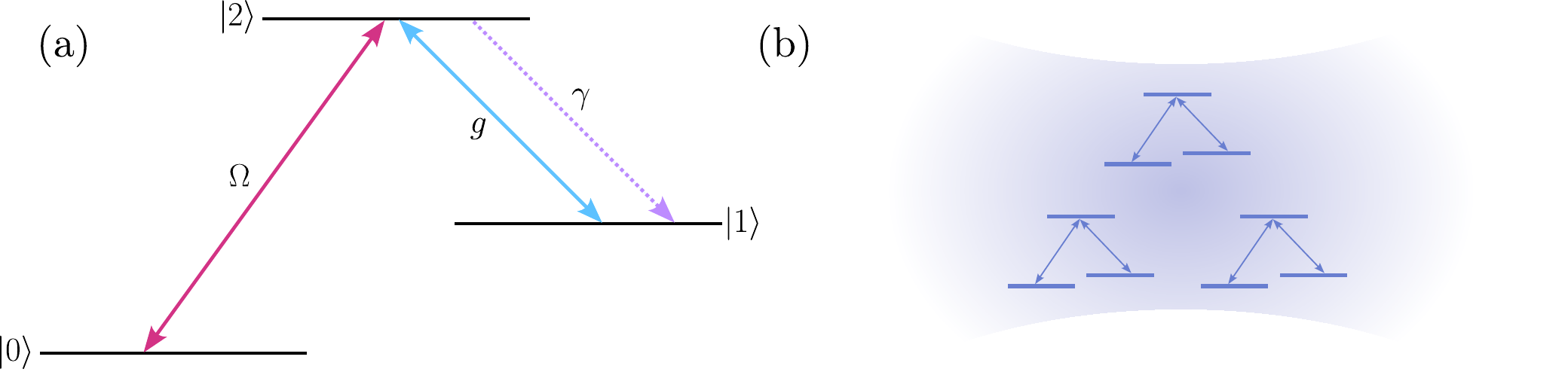}
\caption{(a) Diagram of the energy levels of the considered atom. $N$ such atoms form a quantum battery. (b) The quantum battery placed in the charger, i.e., in the optical cavity. The cavity field mode is initially in a vacuum state. The energy is supplied via a classical laser field.}
\label{fig:schemat_adia}
\end{figure}

In our proposal, the quantum battery comprises identical $N$ atoms. Each atom is modelled by the $\Lambda$ configuration with the ground state $|0\rangle$, the excited state $|2\rangle$ and the long-living metastable state $|1\rangle$ as shown in Fig.\ref{fig:schemat_adia}.
Unlike the usual $\Lambda$-type system, we assume that in the $\Lambda$-type system considered here, the excited state $|2\rangle$ decays to the metastable state $|1\rangle$ only, which means that there is no dipole-allowed transition between levels $|0\rangle$ and $|2\rangle$. We will later discuss how to emulate such a system using a ${}^{87}\mathrm{Rb}$ atom. The QB is described by the Hamiltonian $H_B=\sum_{j=1}^{N} (E_1 \sigma_{11}^{(j)} + E_2 \sigma_{22}^{(j)})$, where we assumed $E_0=0$ and introduced flip operators $\sigma^{(j)}_{mn}=\ket{m^{(j)}}\bra{n^{(j)}}$. 

Although we include both excited energy levels ($\ket{2}$ and $\ket{1}$) in the Hamiltonian $H_B$, we are really only interested in energy that can be stored for long periods of time. The lifetime of the excited state $|2\rangle$ is very short. The excited energy level $\ket{2}$ decays to the long-living metastable level $\ket{1}$ very quickly, releasing energy to the environment, and thus, the expected value $\bracket{H_B}$ has no practical meaning. Likewise, ergotropy is not good figure of merit in this scheme, because it is calculated using $H_B$. On the other hand, the sum of the populations of metastable states $\langle\sum_{j=1}^{n}\sigma_{11}^{(j)}\rangle$ has obvious operational meaning --- it is a number of excitations that can be stored over very long time in QB. Therefore, $\langle\sum_{j=1}^{n}\sigma_{11}^{(j)}\rangle$ plays the role of the primary figure of merit in this scheme instead of the mean energy or the ergotropy. Consequently, we will use the term "charged QB" to refer to the state of the battery in which all its atoms are in a metastable state. Of course, then we will use the term "stored energy" to refer to $E_1\cdot \langle\sum_{j=1}^{n}\sigma_{11}^{(j)}\rangle$. As we will soon see, this charged state of QB is a steady state to which QB asymptotically tends. However, the time of the protocol has to be finite, and therefore, we will also use the term "charged QB", when QB reaches $99\%$ of the asymptotic charge. Then the term "charging time" corresponds to the time of reaching $99\%$ of the asymptotic charge. It is worth to note that the mean energy or the ergotropy has practical meaning for the final state of the charging protocol, because it is close to a steady state.

We assume that QB initially is empty, i.e., all atoms are in their ground state $\ket{\Psi_0}=\bigotimes_{j=1}^{N} |0\rangle_{j}$. In order to charge QB we place it in an optical cavity and illuminate QB by a classical laser field, as shown in the panel (b) in Fig.\ref{fig:schemat_adia}. Initially, the cavity is prepared in the vacuum state $|0\rangle_{\textrm{cav}}$, which is much easier to prepare than the Fock state $|N\rangle_{\textrm{cav}}$ assumed in the usual Dicke battery. The appropriate Hamiltonian that governs the evolution of the QB-charger system during the charging procedure can be written in a rotating frame as:
\begin{eqnarray}
    H & =& g \sum_{j=1}^{N} \big(a^\dagger \sigma_{12}^{(j)} + \sigma_{21}^{(j)} a\big) 
    + \Omega \sum_{j=1}^{N} \big(\sigma_{20}^{(j)} + \sigma_{02}^{(j)}\big)\, ,
    \label{eq:Hhermitian}
\end{eqnarray}
where $a$ denotes the annihilation operator for the cavity mode. External driving with Rabi frequency $\Omega$ excites atomic level $|2\rangle$, and $g$ describes the coupling of the cavity mode $a$ with the transition $|2\rangle\leftrightarrow |1\rangle$, which is the common value for $N$ atoms. 

As our system is an open one, we include the decay of the atomic excited level of the $ j$-th atom via the collapse operator $C_j$ which describes the decay from level $|2\rangle$ to $|1\rangle$, and the decay of the cavity mode --- emission from the cavity,  via the operator $C_0$ as follows:
\begin{eqnarray}
    C_0 &=&\sqrt{\kappa}\,a \, ,\nonumber\\
    C_j &=&\sqrt{\gamma}\,\sigma_{12}^{(j)}\, .
    \label{eq:coll}
\end{eqnarray}
Note that all atoms interact with the cavity field mode collectively (the Tavis-Cummings model), but their decay is modelled as independent local jumps, not a collective superradiant channel.

The damping in the dynamics of a system includes the energy losses and the loss of coherence. In the full quantum description given by the master equation, all of these processes are present and are usually assumed to be the obstacles in the applications of a given quantum system. 

In the present work, we focus on the possible benefits of damping processes induced by unidirectional quantum jumps. In particular, we are interested in benefits, such as the possibility of reducing battery charging time and increasing the amount of extractable energy. For that purpose, we will consider the behaviour and compare three descriptions of the quantum battery dynamics:
\begin{itemize}
    \item[(a)] without damping: by solving the Schr\"{o}dinger equation with a Hermitian Hamiltonian $H$ given by Eq.~(\ref{eq:Hhermitian}),
    \item[(b)] with a coherent part of damping, i.e., in the absence of quantum jumps: by solving the non-unitary Schr\"{o}dinger equation with the non-Hermitian Hamiltonian $H_{\rm nH} = H - \frac{i}{2}\sum_{j=0}^{N} C_j^{\dagger}C_j$,
    \item[(c)] with damping including quantum jumps: full master equation for the system's density matrix.
\end{itemize}

We want to emphasize that the first two descriptions, i.e. the descriptions (a) and (b), are just mathematical models that serve as tools to investigate the influence of damping on the charging protocol and only the description (c) corresponds to a realistic and practical physical scheme. The description (a) is a valid quantum model of the evolution of the isolated system, but in real experiments the system cannot be perfectly isolated from the environment. The description (b) is a valid quantum model of the conditional evolution of the open system during time intervals without any quantum jumps events~\cite{Dalibard92,Hegerfeldt93,PlenioKnight_traj,carmichael_traj}. Physically the description (b) can be realised using post-selection, but charging QB using post-selection is not practical, because one has to repeat the charging protocol many times and reject all runs of the protocol where quantum jump occurs.

\subsection*{Charging protocol}

\begin{figure}[ht]
\centering
\includegraphics[width=0.8\linewidth]{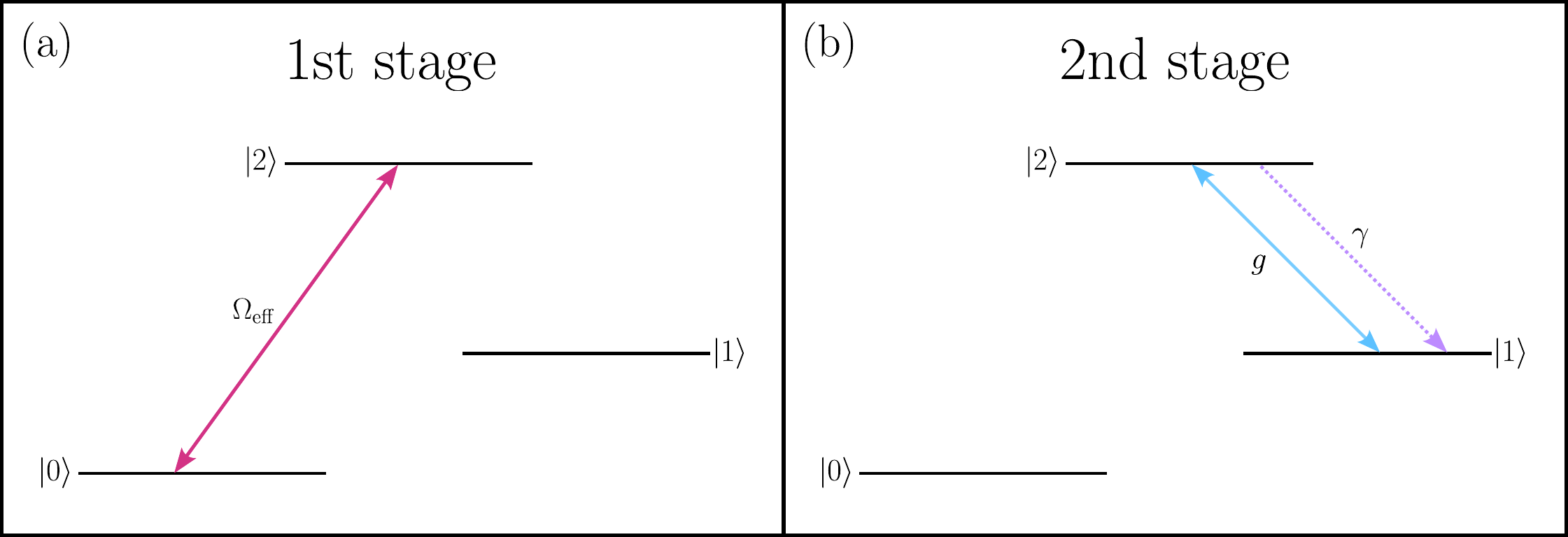}
    \caption{The charging protocol consists of two stages. In the first stage a very intense classical laser field moves the population from the ground state to the excited state. In the second stage the laser is turned off and the population is transferred from the excited state to the metastable state via different processes.}
    \label{fig:Protocol}
\end{figure}

At the beginning of the protocol, all atoms are in their ground state, and the cavity is prepared in the vacuum state. The external field characterised by the effective Rabi frequency populates the atomic level $|2\rangle$, which is later allowed to decay to the state $|1\rangle$. 
We have chosen the atomic system in such a way that there are no other allowed dipole transitions to different atomic levels. We assume that the Rabi frequency is very large, so we can move the entire population between the ground and the excited level $|2\rangle$, which afterwards decays to the state $|1\rangle$. Since the transition $|2\rangle\leftrightarrow |1\rangle$ is a dipole-allowed transition, the atom-cavity coupling can play a significant role, and therefore we use the cavity mode to speed up the process of moving the population via the collective interaction known from the Dicke QB. For the purposes of using the described atomic system in the lambda configuration as a quantum battery, we propose to utilise the following two-step charging protocol, shown schematically in Fig.(\ref{fig:Protocol}). 

\subsubsection*{First stage of the protocol}
In the first stage of the protocol, we turn the laser on to transfer a population from the ground state $|0\rangle$ to the excited state $|2\rangle$ in all atoms.
We assume that the laser pulse is very intensive, i.e. $\Omega$ is much greater than all other parameters characterising the QB-charger system. Therefore, the Hamiltonian can be well approximated by the Rabi model
\begin{eqnarray}
    H_1 &\approx&\Omega \sum_{j=1}^{N} \big(\sigma_{20}^{(j)} + \sigma_{02}^{(j)}\big)\, .
    \label{eq:H1}
\end{eqnarray}
The population is transferred to the excited state after a period of time $t_1=\pi/\Omega$, and then the laser is turned off. Since we assume that $\Omega\gg\gamma$ we neglect dissipation in this stage. So, at the end of this stage, the state of the joint QB-charger system is given by 
\begin{eqnarray}
\label{eq:Phi1}
\ket{\Phi_1}&=&(\bigotimes_{j=1}^{N} |2\rangle_{j})\bigotimes |0\rangle_{\textrm{cav}}\, .
\end{eqnarray}

\subsubsection*{Second stage of the protocol}
In the second stage of the protocol, the dissipation of the system is considered in two different scenarios. This stage of the protocol is more interesting than the previous one because now decay can play a constructive role. Decay can be useful for moving the population from the excited state $|2\rangle$ to the metastable state $|1\rangle$. Moreover, the atomic transition between states  $|1\rangle$ and $|2\rangle$ is also coupled to the cavity mode $a$ to exploit collective effects. In this stage, the laser is turned off ($\Omega=0$), and therefore the Hamiltonian takes the form
\begin{eqnarray}
\label{eq:H2}
    H_2 & =&g \sum_{j=1}^{N} \big(a^\dagger \sigma_{12}^{(j)} + \sigma_{21}^{(j)} a\big)\, . 
\end{eqnarray}
The evolution in this stage differs significantly depending on which model we use: (a) without damping, (b) only with a coherent part of damping, or (c) full damping model including quantum jumps. We will investigate numerically the effect of damping and quantum jumps on population transfer during the second stage of the protocol in the next subsection.

\subsection*{Numerical investigations of the effect of quantum jumps}

The main objective of our considerations is to show the beneficial influence of the damping process on
the amount of extractable energy stored in the QB and the charging speed. In particular, we are going to discuss the influence of quantum jumps on battery evolution during the second stage of our charging protocol. For that analysis, we will discuss the damping within the two approaches: without and with quantum jumps. 

The first approach is based on the theory of quantum trajectories~\cite{Dalibard92,Hegerfeldt93,PlenioKnight_traj,carmichael_traj}. According to this theory, the evolution of a state of an open system under absence of quantum jumps is described by the non-Hermitian Hamiltonian in the form of
\begin{eqnarray}
  H_{\rm nH} = H_2 - \frac{i}{2}\sum_{i=0}^{N} C_i^{\dagger}C_i\, ,
\end{eqnarray}
which, after expressing the appropriate Hamiltonian $H_2$ and collapse operators, can be written as:
\begin{eqnarray}
\label{eq:HnH}
  H_{\rm nH} = g \sum_{j=1}^{N} \big(a^\dagger \sigma_{12}^{(j)} + \sigma_{21}^{(j)} a\big)
  - i\frac{\gamma}{2}\sum_{j=1}^{N}\sigma_{22}^{(j)} - i\frac{\kappa}{2} a^\dagger a \, .
\end{eqnarray}
The non-unitary evolution is not trace-preserving and requires renormalising of the state during time evolution. Since the initial state in the second stage~(\ref{eq:Phi1}) is a pure state, the solution of the non-unitary Schr\"{o}dinger equation with the non-Hermitian Hamiltonian~(\ref{eq:HnH}) is also a pure state. Therefore, this approach includes only the coherent part of the damping. 

The second approach uses the master equation for the statistical operator $\rho$. In the master equation approach, individual collapse events are not explicitly tracked. Although the solution inherently incorporates the effects of these events on the dynamics of the system, it does not provide information on the precise times at which they occur. As a result, although the overall influence of quantum jumps is captured statistically, the exact trajectory of each jump remains inaccessible, reflecting the fundamental stochastic nature of dissipation and decoherence in open quantum systems. The master equation that describes the evolution of the QB-charger system during the second stage of the charging protocol is given by
\begin{eqnarray}    
\label{eq:master}
\dot{\rho}=-i [H_2,\rho]
+\frac{1}{2}\sum_{i=0}^{N} \left(2 C_{i}\rho C^{+}_{i} 
- C^{+}_{i} C_{i}\rho - \rho C^{+}_{i} C_{i}\right) \, .
\end{eqnarray}
In the equation given in (\ref{eq:master}), the quantum-jump terms are also included. Therefore, the evolution governed by the master equation captures both processes present in an open quantum system dynamics: a coherent part of dissipation and sudden changes in the wave function introducing decoherence.

In order to make the role of atomic damping clearly visible, let us assume the weak coupling regime, i.e., $g<\gamma$, and let us assume that $\kappa=0$. Specifically, we set the following parameters: $(g\, , \gamma\, , \kappa)=2\pi\times (1, 6, 0)$~MHz in our computations.
\begin{figure}[ht]
\centering
\includegraphics[width=0.8\linewidth]{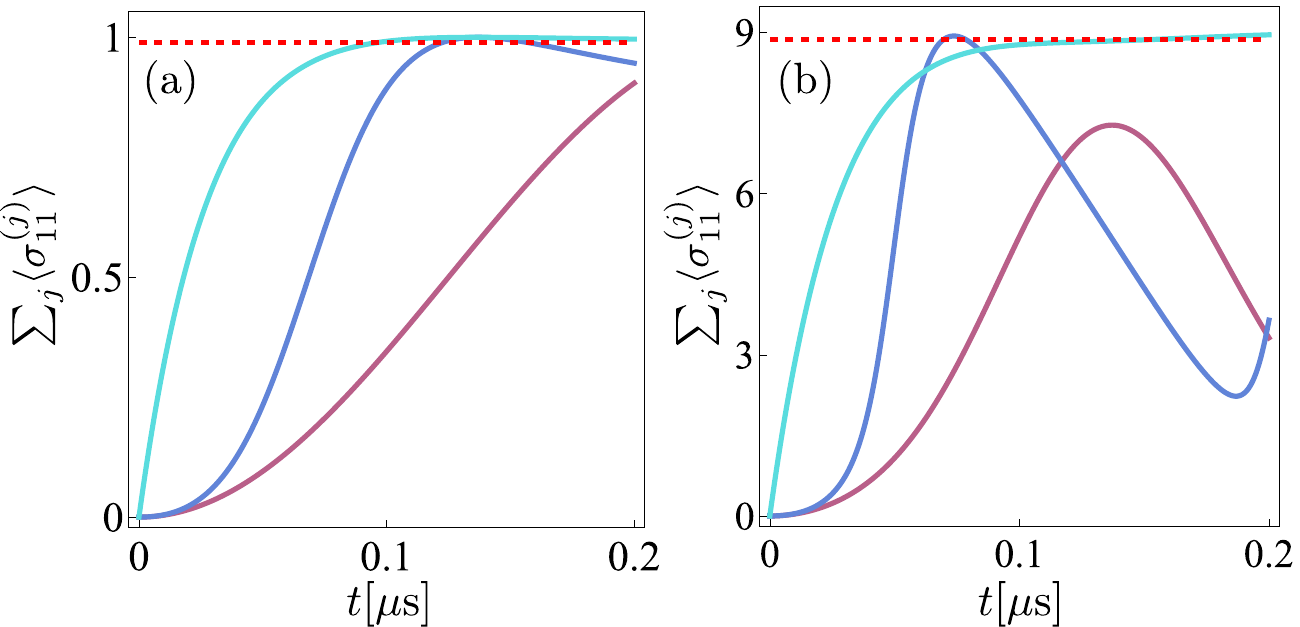}
    \caption{The sum of populations of the metastable states calculated numerically using the Hermitian Hamiltonian~(\ref{eq:H2}) (pink), the non-Hermitian Hamiltonian~(\ref{eq:HnH}) (blue) and the master equation~(\ref{eq:master}) (cyan) as a function of time for parameters $(g\, , \gamma\, , \kappa)=2\pi\times (1, 6, 0)$~MHz. In the panel (a) the quantum battery is composed of one atom, and in the panel (b) the quantum battery is composed of nine atoms. On both panels the red dashed line indicates 99\% of the maximum population of metastable state reached in the steady state.}
    \label{fig:N1N9}
\end{figure}
In Figure~\ref{fig:N1N9}, one can see the comparison of the three models mentioned above, where we plot the sum of the populations of metastable states $\langle\sum_{j=1}^{n}\sigma_{11}^{(j)}\rangle$ as a function of time.
In panel (a), we present results obtained assuming that QB is composed of one atom ($n=1$), and therefore that QB is fully charged in this case when the sum reaches the value one. In panel (b), QB consists of nine atoms ($n=9$), so it is fully charged when the sum reaches nine. In Figure~\ref{fig:N1N9} it is obvious that dissipation is beneficial for the battery charging process. Both the coherent part of dissipation and quantum jumps shorten the time to reach the maximum battery charge. In panel (a), the sum reaches one much faster for the non-Hermitian Hamiltonian~(\ref{eq:HnH}) than for the Hermitian one~(\ref{eq:H2}). However, QB reaches a fully charged state even faster when we also take quantum jumps into account. In this weak coupling regime, there is no oscillatory behaviour in solution of the master equation~(\ref{eq:master}) --- QB charge level asymptotically tends to the value corresponding to a fully charged battery. One can easily check that the pure state $\rho_{ss}=\ket{\Phi_{ss}}\bra{\Phi_{ss}}$, where
\begin{eqnarray}
\label{eq:SteadyState}
\ket{\Phi_{ss}}&=&(\bigotimes_{j=1}^{N} |1\rangle_{j})\bigotimes |0\rangle_{\textrm{cav}}
\end{eqnarray}
is a steady state solution of the master equation~(\ref{eq:master}). This means that all energy stored in this state is extractable. The ergotropy is equal to the energy stored in this steady state despite the fact that dissipation, which usually leads to decoherence, is fully taken into account. This is due to the unidirectional quantum jumps to the metastable state. Thus, unidirectional quantum jumps not only reduce the duration of the second stage of the charging process but also enable the quantum battery to become fully charged, with all stored energy being extractable. Naturally, this improvement comes at the cost of reduced charging efficiency. During each quantum jump $|2\rangle\to |1\rangle$ , energy $E_2 - E_1$ is radiated into the environment and irretrievably lost.

In panel (b) of Figure~\ref{fig:N1N9}, we have investigated the effect of quantum jumps on the charging process in the case where QB consists of nine atoms. Comparing both panels, it can be seen that the advantage in charging speed provided by quantum jumps compared to that of the system without damping decreases with an increasing number of atoms. Nevertheless, unidirectional quantum jumps continue to play a constructive role, primarily by shortening the duration of the second stage of the charging process. Beyond this temporal advantage, they also improve the battery’s performance in another crucial way. In the absence of damping, the quantum battery is unable to reach a maximum of nine excitations, limiting the amount of energy it can store. However, the presence of unidirectional decay enables the system to overcome this limitation, allowing the battery to achieve a fully excited state and thereby store the maximum extractable energy.

\begin{figure}[ht]
\centering
\includegraphics[width=0.8\linewidth]{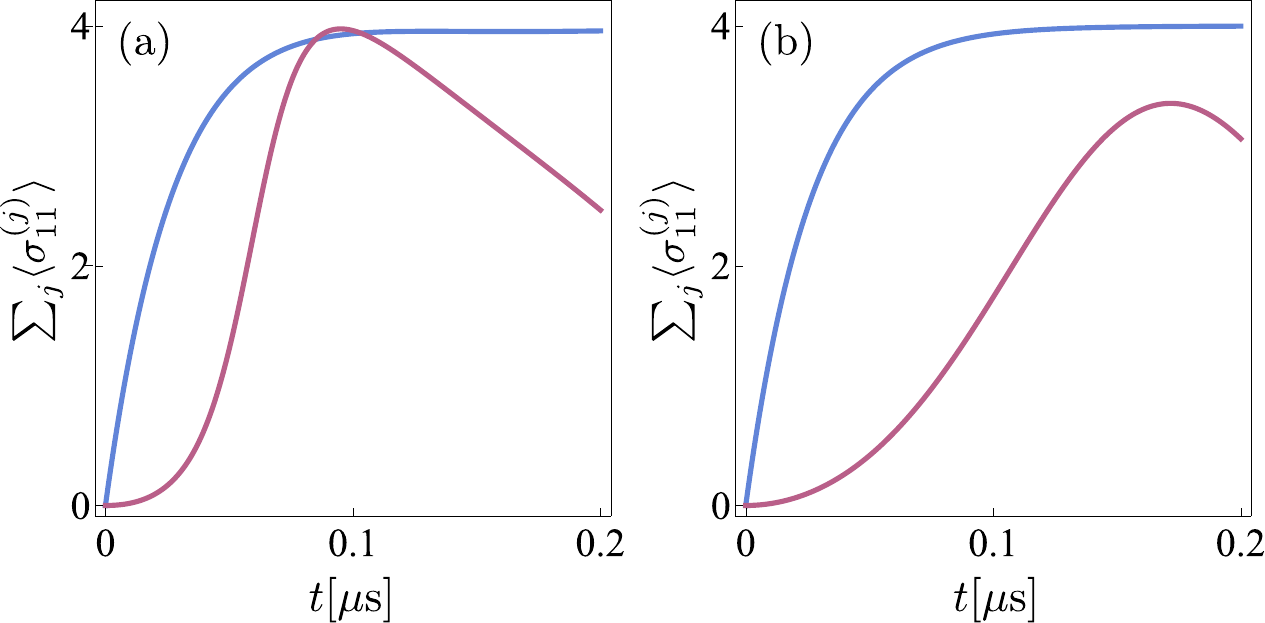}
    \caption{The sum of populations of the metastable states calculated numerically using the non-Hermitian Hamiltonian~(\ref{eq:HnH}) (pink) and the master equation~(\ref{eq:master}) (blue) as a function of time for $N=4$. In the panel (a) for parameters $(g\, , \gamma\, , \kappa)=2\pi\times (1, 6, 0)$~MHz, and in the panel (b) for parameters $(g\, , \gamma\, , \kappa)=2\pi\times (1, 6, 6)$~MHz.}
    \label{fig:N4kappa}
\end{figure}

Now, we want to show the effect of the cavity decay $\kappa$ on the charging process. In Fig.~\ref{fig:N4kappa}, we numerically compare the charging process for a QB consisting of four atoms, described by non-Hermitian Hamiltonian (pink line) and master equation (blue line), for different values of $\kappa$. In panel (b), where the value of $\kappa$ is non-zero, we can observe a significant deterioration in the charging time and the maximum population of the metastable state for the non-Hermitian model. However, the master equation model demonstrated robustness against non-zero value of $\kappa$. It appears that the cavity decay has no effect on either the charging time or the maximum population on the metastable state. This is another advantage of a model that incorporates quantum jumps.

Lastly, let us investigate the role of the cavity in the charging process during the second stage of the protocol. From Fig.~\ref{fig:Protocol}~(b) one can see that the population is transferred from the excited state $\ket{2}$ to the metastable state $\ket{1}$ via two processes: (i) a coherent interaction mediated by the cavity described by the coupling strength $g$, and (ii) a dissipative interaction described by the decay rate $\gamma$. Only the second process leads to the super-extensive charging power scaling~\cite{Colloquium}. We can shorten the time of the second stage of the protocol below $\gamma^{-1}$ only using collective effects present in the Tavis-Cummings model, and therefore, the cavity is necessary. However, the collective effects can be visible only when the condition $g\,\sqrt{N}>\gamma$ is fulfilled. So far we have used the parameters for which the dissipative interaction dominate over the coherent interaction. Let us now show the case, where the coherent interaction dominate. We cannot increase $N$ due to limitations of the numerical methods that we have used. Thus we will increase $g$.
\begin{figure}[ht]
    \centering
    \includegraphics[width=0.9\linewidth]{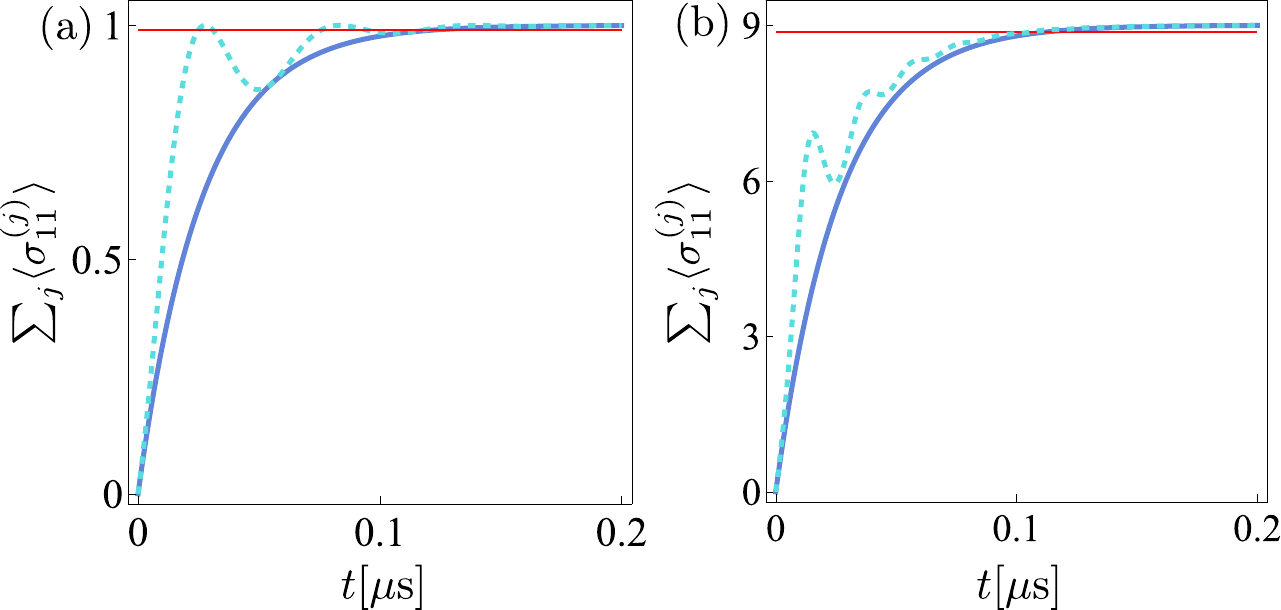}
    \caption{Same as Fig.~\ref{fig:N1N9}, but only solutions of the master equation~(\ref{eq:master}) are shown for $(\gamma\, , \kappa)=2\pi\times (6, 6)$~MHz. In each panel a comparison is presented between $g/2\pi=9$~MHz (cyan) and $g=0$~MHz (blue).}
    \label{fig:g0g9}
\end{figure}

One can see from Fig.~\ref{fig:g0g9} that for $g\,\sqrt{N}>\gamma$ it is possible to charge QB much faster than using only dissipation. This is especially clearly seen in the case of $N=1$. In principle, we can turn off the interaction with the cavity at the first maximum ending the whole protocol. In the case $N=9$, the first maximum does not reach $99\%$ of the asymptotic charge. Nevertheless, we can turn off the interaction with the cavity at the first maximum and then we can let QB asymptotically tend to the steady state.

\subsection*{Physical implementation of the unidirectional quantum jump system}
As was shown in previous sections, unidirectional quantum jumps lead to very interesting and useful phenomena. Therefore, let us now investigate whether it is possible to realise such quantum jumps in a real physical system. It is not possible to observe the behaviour presented in Fig.~\ref{fig:schemat_adia} (a) in a real three-level atom, but it is possible to find a real atom with more than three levels, which is efficiently modelled by the three-level atom with unidirectional quantum jumps. Here, we propose to use a ${}^{87}\mathrm{Rb}$ atom trapped inside an optical cavity to realise such a quantum system. We will use levels $|5S_{1/2}, F=1, m_{F}=1\rangle$, $|5S_{1/2}, F=2, m_{F}=2\rangle$, $|5P_{3/2}, F=3, m_{F}=3\rangle$, $|4D_{3/2}, F=3, m_{F}=3\rangle$, and $|5P_{1/2}, F=2, m_{F}=2\rangle$ and we will denote them as $|0\rangle$, $|1\rangle$, $|2\rangle$, $|3\rangle$, and $|4\rangle$, respectively. 
\begin{figure}[ht]
    \centering
    \includegraphics[width=1.0\linewidth]{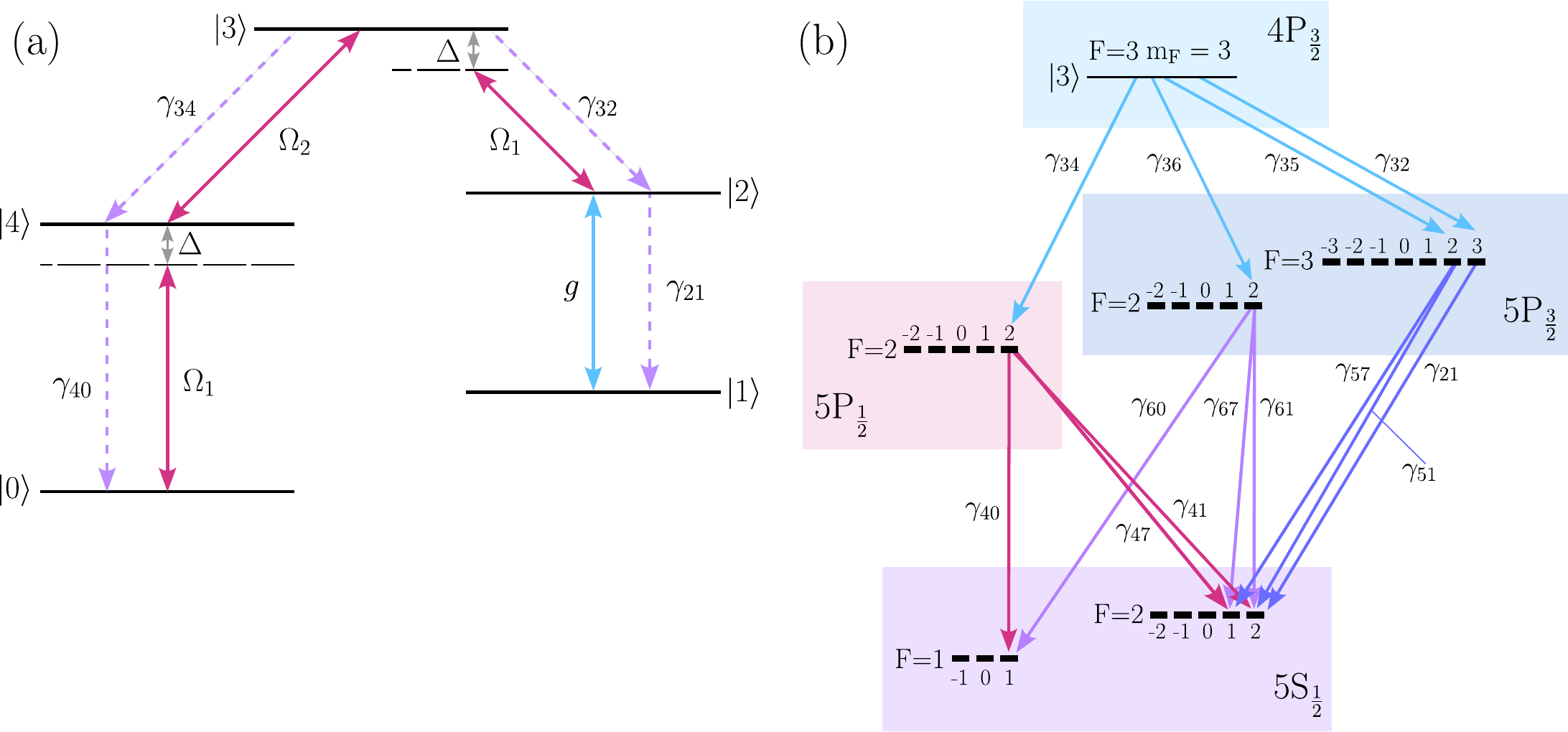}
    \caption{(a) Level scheme which can be realized using a ${}^{87}\mathrm{Rb}$ atom. Providing that the levels $\ket{3}$ and $\ket{4}$ can be adiabatically eliminated this quantum system works efficiently in the charging protocol like the $\Lambda$-type system presented in Fig.~\ref{fig:schemat_adia}. (b) All possible quantum jumps that can occur during the charging protocol.}
    \label{fig:Rb87}
\end{figure}
The level scheme is shown in Fig.~\ref{fig:Rb87} (a). As one can see from this figure the transitions $\ket{0}\leftrightarrow\ket{4}$, $\ket{4}\leftrightarrow\ket{3}$ and $\ket{3}\leftrightarrow\ket{2}$ are driven by classical laser fields with the coupling strengths $\Omega_1$, $\Omega_2$ and $\Omega_1$, respectively. The polarizations of these laser fields are as follows: $\sigma^+$, $\sigma^+$, and $\pi$, respectively. Both laser fields $\Omega_1$ are detuned from the corresponding transitions by $\Delta$. The transition $\ket{2}\leftrightarrow\ket{1}$ is coupled to the optical cavity mode with the coupling strength $g$. Only these four transitions are allowed. All other transitions are forbidden because of selection rules or large detunings. The excited states $\ket{2}$, $\ket{3}$, and $\ket{4}$ spontaneously decay with the rates $\gamma_2/2\pi=6$~MHz, $\gamma_3/2\pi=1.85$~MHz, and $\gamma_4/2\pi=5.7$~MHz, respectively. The metastable state $\ket{1}$ does not experience decay, and therefore energy can be stored in this level for a very long time. The coherent part of dynamics is confined to the subspace spanned by $\ket{0}$, $\ket{1}$, $\ket{2}$, $\ket{3}$, and $\ket{4}$, but some quantum jumps can move the population outside the subspace. The atom can spontaneously jump from the excited state $\ket{3}$ to the states $\ket{2}$ and $\ket{4}$ but also to the states $|5\rangle\equiv |5P_{3/2}, F=3, m_{F}=2\rangle$ and $|6\rangle\equiv |5P_{3/2}, F=2, m_{F}=2\rangle$ as is illustrated in Fig.~\ref{fig:Rb87} (b). Likewise, the atom can jump from the states $\ket{2}$, $\ket{4}$, $\ket{5}$, and $\ket{6}$ to the states $\ket{0}$, $\ket{1}$, and $\ket{7}\equiv |5S_{1/2}, F=2, m_{F}=1\rangle$. For the ${}^{87}\mathrm{Rb}$ atom, transition rates that correspond to these jumps are given by:
\begin{equation}\label{eq:Gammas} 
    \begin{gathered}
    (\gamma{21},,\gamma{32},,\gamma{34},,\gamma{35},,\gamma{36},,\gamma{40},,\gamma{41},,\gamma{47},,\gamma{51},,\gamma{57},,\gamma{60},,\gamma{61},,\gamma_{67})/2\pi =\\
    (6,, 0.72,, 0.65,, 0.24,, 0.24,, 2.5,, 1.6,, 1.6,, 2,, 4,, 3,, 2,, 1),\mathrm{MHz}, .
    \end{gathered} 
\end{equation}

Of course, $\gamma_{2}=\gamma_{21}$, $\gamma_{3}=\sum_i \gamma_{3i}$ and $\gamma_{4}=\sum_i \gamma_{4i}$. The desired quantum jump, i.e. $|2\rangle\to |1\rangle$, is the most probable, because $\gamma_{21}$ is bigger than other decay rates and the mean population of the state $\ket{2}$ is much bigger than mean population of other excited states due to detuning. All other quantum jumps are unwanted, and thus, we will select parameter values such that the probabilities of these jumps remain sufficiently small to be safely neglected. Therefore, and for clarity, levels $\ket{5}$, $\ket{6}$ and $\ket{7}$ are not present in the diagram shown in Fig.~\ref{fig:Rb87} (a).

The evolution of the system presented above is governed by the Hamiltonian, which in the rotating frame is given by
\begin{eqnarray}
\label{eq:Hamiltonian_Rb}
    H& =&\Delta (\sigma_{33}+\sigma_{44}) + \Omega_1 (\sigma_{04} + \sigma_{32}+ \mathrm{H.c})+ \Omega_2 (\sigma_{34} + \mathrm{H.c}) +g (a^\dagger \sigma_{12} + \mathrm{H.c})\, .
\end{eqnarray}
From Fig.~\ref{fig:Rb87} (a) one can see that if $\Delta$ is large compared to $\Omega_1$ and $\Omega_2$, then the populations of excited states $\ket{3}$ and $\ket{4}$ are negligibly small. In such a case, we can adiabatically eliminate these two excited states and use in calculations an effective Hamiltonian
\begin{eqnarray}
\label{eq:Hamiltonian_Rb_eff}
    H^{\prime} & =&\Omega_{\mathrm{eff}} (\sigma_{20} + \sigma_{02}) + g (a^\dagger \sigma_{12} + \mathrm{H.c})\, ,
\end{eqnarray}
where
\begin{eqnarray}
\label{eq:Omega_eff}
    \Omega_{\rm eff}&=&\frac{1}{4}\Big(2\Omega_2 - \sqrt{(\Delta+\Omega_2)^2+4\Omega_1^2} + \sqrt{(\Delta-\Omega_2)^2+4\Omega_1^2}\Big)\, ,
\end{eqnarray}
The exact condition which must be satisfied to make the effective Hamiltonian $H_{\mathrm{eff}}$ work properly is given by
\begin{eqnarray}
\label{eq:condition_adia}
    \frac{|\Omega_1|}{\sqrt{2}(|\Delta|-|\Omega_2|)} &\ll& 1
\end{eqnarray}

In order to make the probability of no quantum jumps during the first stage of the protocol close to one, we also require $\gamma_2,\, \gamma_3,\, \gamma_4 \ll \Omega_{\rm eff}$. In order to realise the dynamics shown in Fig.~\ref{fig:Protocol} (a), we also assume that $g\ll \Omega_{\rm eff}$. One can check that all the aforementioned conditions are satisfied for the parameter regime:
\begin{eqnarray}
\label{eq:ParametersRegime}
    (\Delta ,\,\Omega_1,\,\Omega_2,\, g, \,\gamma_2,\,\gamma_3,\,\gamma_4,\,\kappa)/2\pi &=& (10000,\, 1300,\, 4000,\, 10,\, 6,\, 1.85,\, 5.7,\, 4)\,\mathrm{MHz}\, .
\end{eqnarray}
All of these parameters should be feasible with current technology. We set the value of the cavity decay rate, which was achieved in the experiment reported by Hamsen \emph{et al.}~\cite{Hamsen17PRL}. The atom-cavity coupling constant $g/2\pi=10$~MHz is even smaller than the value achieved in the experiment~\cite{Hamsen17PRL}. The detuning $\Delta$ is very large, but it does not lead to drive other atomic transitions than those shown in Fig.~\ref{fig:Rb87} (a) due to selection rules. 

Let us now test the validity of the effective model and calculate the probability of no jump during the first stage. To this end, we will use the non-Hermitian Hamiltonians
\begin{eqnarray}
\label{eq:HRbnH}
    H_{\textrm{nH}}&=& H 
    -i \sum_{i=2}^{4}\frac{\gamma_i}{2}\sigma_{ii}-i \frac{\kappa}{2} a^{\dagger} a\, ,\\
    \label{eq:HRbnH'}
    H_{\textrm{nH}}^{\prime} & =& H^{\prime}-i\frac{\gamma_2}
    {2}\sigma_{22}-i \frac{\kappa}{2} a^{\dagger} a\, .
\end{eqnarray}
Using these two non-Hermitian Hamiltonians and the parameter values given by~(\ref{eq:ParametersRegime}) we can check if the transfer of the population from the ground state to the excited state $\ket{2}$, which is crucial in the first stage of the charging protocol, is properly described in the effective model. 
\begin{figure}[ht]
\centering
\includegraphics[width=0.8\linewidth]{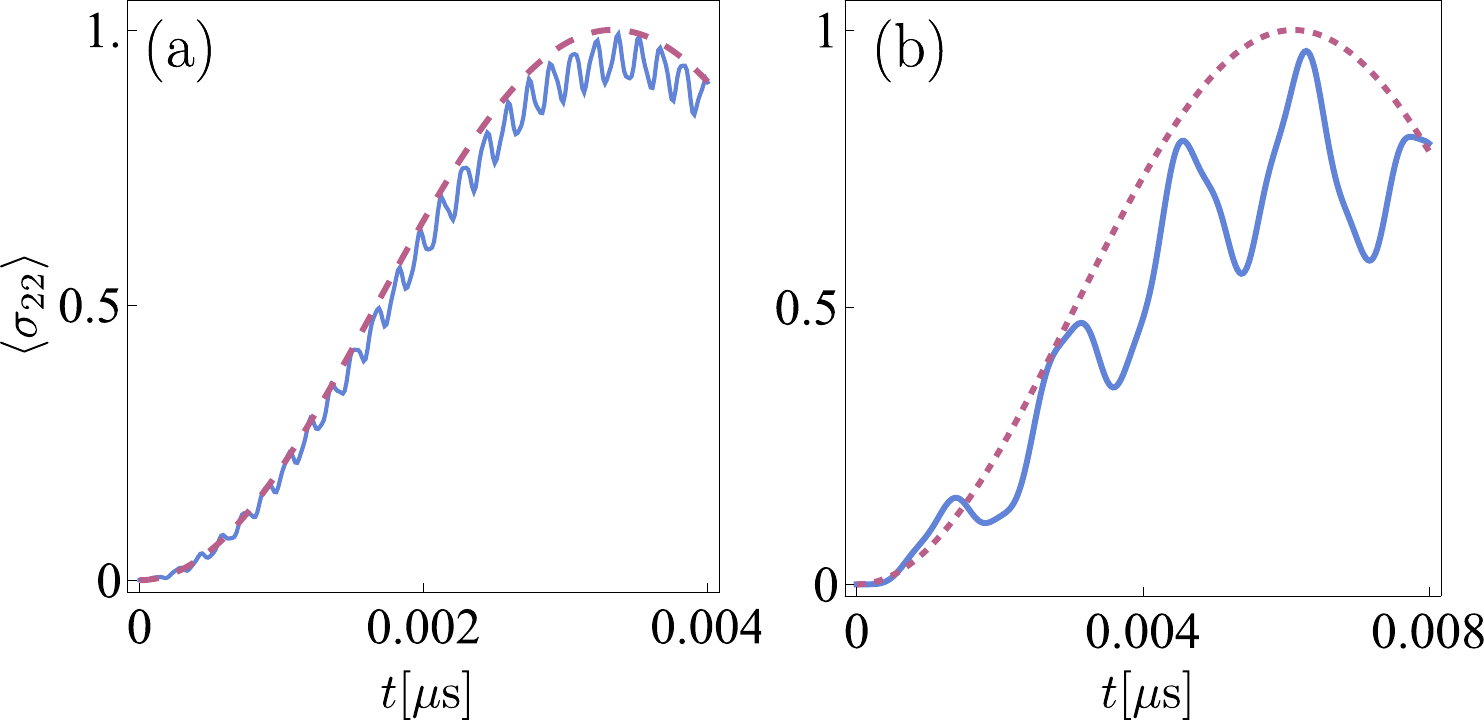}
\caption{(a) The population of the excited state $\ket{2}$ as a function of time during the first stage of the protocol calculated using the general Hamiltonian~(\ref{eq:HRbnH}) (solid curve) and the effective Hamiltonian~(\ref{eq:HRbnH'}) (dashed curve). Both the models, i.e. the general and the effective, are compared using the parameter regime given in Eq.~(\ref{eq:ParametersRegime}). (b) Same as panel (a), but for the parameters regime given in Eq.~(\ref{eq:ParametersRegime2}). One can see that the parameter regime~(\ref{eq:ParametersRegime}) better justifies adiabatic elimination than~(\ref{eq:ParametersRegime2}).}
    \label{fig:GenVsEff}
\end{figure}
From Fig.~\ref{fig:GenVsEff} (a) it is seen that the results obtained using the effective model are in good agreement with the results obtained from the general model. One can also see that it is possible to transfer the population from the ground state to the excited state $\ket{2}$ almost perfectly. If the lasers are turned on for the time $t=3.37$~ns the population of the excited state $\ket{2}$ is equal to $0.992$ and the probability that no quantum jump occurs during this time is equal to $0.938$. Keep in mind that not all quantum jumps are obstacles during the first stage. The quantum jump $\ket{2}\rightarrow\ket{1}$ is even desired.

Let us now investigate the effect of detuning $\Delta$ on the time of the first stage. A large detuning ($\Delta\gg\Omega_1$ and $\Delta\gg\Omega_2$) is necessary to suppress the population of the excited states $\ket{3}$ and $\ket{4}$, and therefore suppresses unwanted loss. The bigger $\Delta$, the smaller the populations of these excited states. On the other hand, the time of the first stage of the protocol also depends on $\Delta$, and, of course, we want this time to be as short as possible. To be more precise, the time of the first stage depends on the effective Rabi frequency $\Omega_{\rm eff}$, and we want $\Omega_{\rm eff}$ to be as large as possible. One can see that it might be difficult to reconcile these two conditions, i.e., there is some kind of trade-off. To gain a deeper insight into the trade-off, let us introduce two quantities $\varepsilon_1=\Omega_1/\Delta$ and $\varepsilon_2=\Omega_2/\Delta$. These two quantities solely determine how well the Hamiltonian~(\ref{eq:Hamiltonian_Rb_eff}) approximates the general Hamiltonian~(\ref{eq:Hamiltonian_Rb}), because we can rewrite the condition~(\ref{eq:condition_adia}) to the form
\begin{eqnarray}
\label{eq:condition_adia2}
    \frac{|\varepsilon_1|}{\sqrt{2}(1-|\varepsilon_2|)} &\ll& 1 \, .
\end{eqnarray}
Using these quantities, one can roughly approximate the effective Rabi frequency by $\Omega_{\rm eff}\approx \Delta\,\varepsilon_1^2\,\varepsilon_2$. Provided that the optimal values of $\varepsilon_1$ and $\varepsilon_2$ are known, we can consider two simple methods to increase $\Omega_{\rm eff}$. In the first method we keep $\Delta$ constant, while we increase $\varepsilon_1$ and $\varepsilon_2$ by increasing $\Omega_1$ and $\Omega_2$. In this case the scheme has a clear trade-off. We can increase $\Omega_{\rm eff}$ but at the expense of increasing the left hand side of the inequality~(\ref{eq:condition_adia2}). Then the time of the first stage will be shorter but the population of the excited states $\ket{3}$ and $\ket{4}$ will be bigger leading to bigger unwanted losses. In the second method we keep $\varepsilon_1$ and $\varepsilon_2$ constant, while increasing $\Delta$. In this case there is no trade-off. We increase $\Omega_{\rm eff}$ without changing the value of the left hand side of the inequality~(\ref{eq:condition_adia2}). Therefore, the second method is beneficial over the first one.

Unfortunately, the problem of finding optimal values of $\Delta$, $\Omega_1$ and $\Omega_2$ is difficult in the general model described by the Hamiltonian~(\ref{eq:Hamiltonian_Rb}) and can be solved only numerically. The difficulty comes from the rapid oscillations, which one can see in both panels in Fig.~\ref{fig:GenVsEff}. At the end of the first stage these fast oscillations have to end at the maximum value. Therefore, it is necessary to tune these fast oscillations properly~\cite{finetuning1,finetuning2}. To this end we start by finding optimal values of $\Delta$, $\varepsilon_1$ and $\varepsilon_2$ using the effective model, and then we use those values as a starting point in a numerical searching procedure. 

In Fig.~\ref{fig:GenVsEff} (b) one can see the population of the state $\ket{2}$ as a function of time calculated for the parameter regime in which $\Delta$ is one order of magnitude smaller than in~(\ref{eq:ParametersRegime}):
\begin{eqnarray}
\label{eq:ParametersRegime2}
    (\Delta ,\,\Omega_1,\,\Omega_2,\, g, \,\gamma_2,\,\gamma_3,\,\gamma_4,\,\kappa)/2\pi &=& (950,\, 250,\, 550,\, 10,\, 6,\, 1.85,\, 5.7,\, 4)\,\mathrm{MHz}\, .
\end{eqnarray}
Comparing both panels of Fig.~\ref{fig:GenVsEff} one can see that even big values of $\varepsilon_1$ and $\varepsilon_2$ (i.e. $\varepsilon_1=0.26$ and $\varepsilon_2=0.58$) cannot compensate for such a small $\Delta$, and therefore the time of the first stage of the protocol is longer in panel (b) than in panel (a). So, the scheme requires rather large detuning values. 

One can see that the time evolution of a ${}^{87}\mathrm{Rb}$ atom trapped inside an optical cavity can simulate the time evolution of the quantum system presented in Fig.~\ref{fig:Protocol} (a) during the first step of the charging protocol. It is also obvious that if all lasers are turned off ($\Omega_1=\Omega_2=0$), then the dynamics of the ${}^{87}\mathrm{Rb}$ atom-cavity system is exactly the same as the dynamics of the quantum system presented in Fig.~\ref{fig:Protocol} (b) during the second step, because then (assuming negligible populations of the excited states $\ket{3}$ and $\ket{4}$) the Hamiltonian~(\ref{eq:Hamiltonian_Rb}) becomes 
\begin{eqnarray}
    H& =&g (a^\dagger \sigma_{12} + \mathrm{H.c})\, .
\end{eqnarray}
Moreover, under these conditions during the second stage only the quantum jump $\ket{2}\rightarrow\ket{1}$ is allowed, and therefore a unidirectional quantum jump is present in this real physical system.

We conclude this section by investigating the performance of the charging protocol, when a single ${}^{87}\mathrm{Rb}$ atom plays the role of a quantum battery. This time we take into account all possible quantum jumps presented in Fig.~\ref{fig:Rb87} (b). To this end, we will use the master equation
\begin{eqnarray}    
\label{eq:masterRb}
\dot{\rho}=-i [H,\rho]
+\frac{1}{2}\sum_{i=0}^{N} \left(2 C_{i}\rho C^{+}_{i} 
- C^{+}_{i} C_{i}\rho - \rho C^{+}_{i} C_{i}\right) \, ,
\end{eqnarray}
where $H$ is the general Hamiltonian~(\ref{eq:Hamiltonian_Rb}), $C_0=\sqrt{\kappa}\,a$ and atomic jump operators $C_j=\sqrt{\gamma_{mn}}\,\sigma_{nm}$ correspond to transition rates listed in Eq.~(\ref{eq:Gammas}) and indicated in Fig.~\ref{fig:Rb87} (b). 

For the parameter regime given by~(\ref{eq:ParametersRegime}) we set the time of the first stage $t_1=3.368$~ns, and the time of the second stage $t_2=0.4\,\mu{\rm{s}}$. At the end of the charging protocol the state of the quantum battery can be very well approximated by
\begin{eqnarray}
    \rho_{\rm{QB}}& =& 0.99772\, |1\rangle\langle 1|+ 0.00114\, |0\rangle\langle 0|+0.00114\, |7\rangle\langle 7|\, .
\end{eqnarray}
The frequency splitting between the hyperfine energy levels $5S_{1/2}, F=1$ and $5S_{1/2}, F=2$ is equal to $6.835$~GHz~\cite{steck2001rubidium}, which leads to the energy of levels $\ket{1}$ and $\ket{7}$ equal to $2.827\cdot 10^{-5}$~eV. Therefore, the mean energy stored in QB is equal to $2.824\cdot 10^{-5}$~eV, whereas the ergotropy is equal to $2.817\cdot 10^{-5}$~eV. So, despite taking into account all unwanted losses the ergotropy is still almost equal to the energy stored in QB.

For the parameter regime given by~(\ref{eq:ParametersRegime2}) we set the time of the first stage $t_1=6.3$~ns, and the time of the second stage $t_2=0.5\,\mu{\rm{s}}$. In this case the final state  of the quantum battery can be very well approximated by
\begin{eqnarray}
    \rho_{\rm{QB}}& =& 0.9831\, |1\rangle\langle 1|+ 0.0092\, |0\rangle\langle 0|+0.0077\, |7\rangle\langle 7|\, .
\end{eqnarray}
The mean energy stored in QB is now equal to $2.80\cdot 10^{-5}$~eV, and the ergotropy is now equal to $2.75\cdot 10^{-5}$~eV.

\section*{Discussion}

Quantum batteries commonly studied in the literature are assumed to be an ensemble of two-level atoms. In this paper, we propose the use of $\Lambda$-type three-level atoms with a metastable state instead of two-level atoms to allow the storage of energy over a long time. Previously, a three-level $\Lambda$-type atom was considered as a quantum battery in only a single study~\cite{BelenoNJP24}. Here, we have proposed an unconventional $\Lambda$-type three-level atom, in which the excited state decays exclusively to the metastable state. For such a QB we have presented a two-stage charging protocol and investigated the effect of quantum jumps on the charging process. The numerical results have shown that the unidirectional quantum jumps to the metastable state play a constructive role in QB charging. They shorten the charging time and make it possible to obtain a fully charged QB. Moreover, the unidirectional quantum jumps to the metastable state lead to a steady state which is a pure state, and therefore all energy stored in the QB is fully extractable. These results might be surprising because usually quantum jumps due to randomness of their occurrence introduce decoherence, which decreases amount of energy that can be extracted. However, we have to pay for these benefits with lower charging efficiency because some of the energy taken from the source leaks out to the environment. Finally, we have shown that a ${}^{87}\mathrm{Rb}$ atom can effectively emulate a $\Lambda$-type atom with unidirectional quantum jumps to the metastable state in the two-stage charging protocol. To this end, we have chosen five levels of the rubidium atom such that the ground level is indirectly connected to the metastable level via dipole-allowed transitions, and then we have adiabatically eliminated two auxiliary levels. The parameters that justify the adiabatic elimination should be feasible with current technology.

\section*{Methods}

In our investigations we use the well known quantum trajectory method~\cite{carmichael_traj,PlenioKnight_traj}. Numerical calculations are performed using a python framework for the dynamics of open quantum systems QuTiP~\cite{QuTiP1,QuTiP2}.

\section*{Data availability}
The datasets used and analysed during the current study are available from the corresponding author on reasonable request.

\bibliography{qbbib}

\section*{Acknowledgements}
Scientific work financed from the state funds (Poland) under the program of the Ministry of Education and Science called "Pearls of Science", project number PN/01/0234/2022, founding amount 2540 EUR, total project value 179 674 PLN.

\noindent K.T. and J.P. acknowledge support by the project OP JAC CZ.02.01.01/00/22\_008/0004596 of the Ministry of Education, Youth, and Sports of the Czech Republic.

\noindent K.T. acknowledges funding from the European Union's Horizon Europe research and innovation programme under the Marie Sklodowska-Curie grant agreement No. 101126636.

\section*{Author contributions statement}
E.L., G.C., A.K.-K., K.T. and J.P. conceived the study, E.L. and G.C. performed calculations, all the authors discussed the methods and analysed the results, G.C., E.L. and A.K.-K. wrote the main manuscript text, E.L. created all figures. All the authors were involved in the revision and discussion of the manuscript.

\section*{Additional information}

\textbf{Competing interests} 
The authors declare no competing interests.

\end{document}